# Commonalities: The R.E.A. and High-Speed Rural Internet Access

by

## Laurence J. Malone

**Economics Department**
**Hartwick College**
**Oneonta, NY 13820**

**malonel@hartwick.edu**

**(607) 431 4943**

### ACKNOWLEDGMENTS

Special thanks to Alycia Vivona of the Franklin Delano Roosevelt Library for research assistance on the origins of the Rural Electrification Administration. I have also benefited from conversations with John Willis, Carli Cochi Ficano, and Silvano Wueschner on high-speed Internet access for rural areas like our own. Robert Whaples and an anonymous reviewer for EH.NET raised incisive questions on the history of the R.E.A. Ellen Falduto provided a workspace and a T1 connection at the TeleCenter in Oneonta, New York for my sabbatical. Thanks also to Mark Rhoads and the United States Internet Council for providing a home for an earlier version of this paper. This work is generously supported by a Trustee Research Grant from Hartwick College.

**Introduction**[*]

In 1936, five decades after municipal electrical distribution systems were first constructed in the United States, the process of introducing rural areas to the twentieth-century economy began with the creation of the Rural Electrification Administration. The R.E.A. encouraged rural Americans to form electrical cooperatives to overcome the unwillingness of private utilities to bring power to households, farms and businesses in sparsely populated regions where profits were too low. Today, on the dawn of a new century, a remarkably similar dilemma confronts rural residents—who will provide a broadband infrastructure in places where profit incentives do not exist?

This paper explores commonalities in these two episodes. In the case of the R.E.A., the necessity for an aggressive federal initiative to "wire" rural America, where the market for electricity had failed, is revisited. The absent incentives, which left rural areas literally and figuratively in the dark, are identified and explored. We then examine the similarities between rural electrification and rural high-speed Internet access through how consumers currently and prospectively gain access to broadband Internet service in an incentive-poor market. The regulatory environment created by the Telecommunications Act of 1996 and the Federal Communications Commission over the past decade is considered in light of the failure of the market to deliver high-speed Internet access to rural America. Finally, we pose the question of whether a federal effort equivalent to the R.E.A. is needed to ensure that residents of sparsely populated areas, like their predecessors in the 1930s, are not comparatively disadvantaged in the first decades of the 21st Century.

---

[*] The sections of this paper on the history of the R.E.A. are adapted from Laurence J. Malone, "Rural Electrification Administration." EH.Net Encyclopdia, edited by Robert Whaples, August 15 2001 URL http://www.eh.net/encyclopedia/contents/malone.electrification.administration.rural.php

**Market Failure in Delivering Electricity to Rural Areas Before 1930**

The advent of electricity in the 1880s ushered forward a rapidly expanding domestic market in the United States that would be dominated by heavy industries for a hundred years. The physical scale of the electric utility industry mirrored the national economy that sprung up with it—massive central generating stations, substantial capital investments for network construction, high maintenance costs, and production technologies that were obtrusive and degrading to the natural environment. But the adaptation of electricity to manufacturing and services further liberated the nineteenth-century economy from proximity to moving water and, with welcoming immigration and naturalization policies, accelerated the pace of urbanization.

While urban households gained electricity in large numbers after 1910, the more sparsely populated rural regions of the United States were generally without electricity and were denied the commercial progress it brought. Electrical providers ignored the rural market due to its high network construction costs and the prospect of meager immediate profits. From the supplier standpoint, rural homes, farms and businesses were stretched too far apart and offered too little revenue relative to the cost of investment. Unlike their counterparts in cities, rural residents were expected to advance the financing for the necessary infrastructure to the firm supplying electrical power from a distant location. In rural areas that were serviced, electrical rates in the 1920s were twice as high as urban rates.[1]

The disincentives to investment in electrical infrastructure left rural America increasingly distant from rising living standards in urban and emerging suburban settings. Lacking the greater productive efficiencies secured by the adaptation of electricity, productivity growth in agriculture, the industry that served as the central organizing principle for rural life, lagged other sectors in the economy over the 1880 to 1930 period. Rural demands for the newest manufactured items found in urban American homes—telephones, radios, refrigerators, washing machines, hot water heaters, and household appliances— were latent. Given the widening disparities between rural and urban settings, it was not surprising that rural Americans reverted to the cooperative lifestyles of the nineteenth century as the urban markets for their agricultural products collapsed in the Great Depression.

**The Origins of the New Deal Rural Electrification Initiative**

The failure of the market to deliver affordable electricity led to over thirty state rural power initiatives during the 1920s and early 1930s, as President Herbert Hoover argued that responsibility for rural electrification rested with state government.[2] Governor of New York Franklin Delano Roosevelt aggressively promoted rural electrification, and the New York Power Authority was created in 1931 to develop a substantial new source of inexpensive hydroelectric generating capacity along the St. Lawrence River.[3]  But the Depression led to the collapse of many state power authorities and further raised the bar in discouraging private investment in rural electrical infrastructure. When Roosevelt assumed the Presidency on March 4, 1933, the market for rural electrification no longer existed.

While Roosevelt clearly understood the benefits of rural electrification, it was Morris L. Cooke who provided vision and leadership to rural electrification efforts under the New Deal.  Cooke had led Giant Power, the Pennsylvania rural electrification program, and Roosevelt invited him to address the problem at the federal level.  Using data supplied by the utility industry, electrical engineers, Giant Power, and the U.S. Census of 1930, Cooke authored an eleven-page report in 1934 that provided the foundation for a federal rural electrification program. Cooke provided detailed estimates of the cost per mile of "high wire" distribution lines in rural regions.  He wrote:  "This cost of the line with transformers and meters included for one to three customers will range from $500 to $800 the mile.  To amortize this cost in twenty years at four percent involves a cost to each of the three customers on a mile of line of about one dollar a month."[4]  Studies commissioned by Cooke suggested that household payments for electricity would be a minimum of one dollar per month for the first ten kilowatts of electricity, three cents per kilowatt for the next forty kilowatts, and two cents per kilowatt for the remaining balance.[5]  All told, the estimated cost to provide electricity to 500,000 farms, at an average of three farms per mile of road, was $112 million, or $225 per farm.  In a worst case scenario, if new generating facilities were needed for all 500,000 farms, the 333 power plants required would cost an additional $87 million.  Consequently, Cooke's high-end estimate for the complete electrical infrastructure needed to bring electrical service to 500,000 farms was $200 million, or $400 per farm.[6]  The concluding paragraph of his report states that a new "rural electrification agency" should build the necessary infrastructure since the market would not otherwise furnish electricity to sparsely populated localities.[7]

Presidential Executive Order 7037 created the Rural Electrification Administration, or R.E.A., on May 11, 1935.  With passage of the Norris-Rayburn Act the following year, Congress authorized $410 million in appropriations for a ten-year program to electrify American farms.   The rural cooperative model, which had been successfully employed by Giant Power in Pennsylvania, was adopted by the R.E.A., with Congressional Representatives serving as the administrative liaisons for the formation of cooperatives within their districts.[8] Cooperatives were consumer-owned firms organized to provide electric service to member-customers.  Each cooperative was typically governed by a board of directors elected from the ranks of its customers.  The board established rates and policies, and hired a general manager to conduct the ordinary business of

providing electricity within the service region. Only two restrictions were placed on the formation of cooperatives: they could not compete directly with utility companies, and coop members could not live in areas served by utilities or within a municipality with a population of 1500 or more.[9]

The R.E.A. was essentially a government-financing agency providing subsidized loans for the construction of electrical supply infrastructure in rural regions. The loans were guaranteed by the federal government and had an attractive interest rate and a generous repayment schedule of 25 years. The interest rate initially matched the federal funds rate when the loan was executed, but after 1944 the rate was fixed at 2%.[10] R.E.A. loans furnished the incentive for rural electric cooperatives to form and connect to the existing electrical network at rates comparable to the national average. R.E.A. cooperatives quickly became one of the largest capital investment projects of the New Deal, and low-cost financing for construction of electrical supply infrastructure was the key provision of the program.[11]

## R.E.A.: The Outcomes

The R.E.A. is one of the most immediate and profound successes in the history of federal policy-making. By the end of 1938, after just two years, 350 cooperative projects in 45 states were delivering electricity to 1.5 million farms.[12] The success of the R.E.A. over the long-term was even more impressive, especially as a self-sustained agency. Monies lent through the R.E.A. were largely repaid, as the default rate on loans was less than 1%.[13] By the mid-1950s nearly all American farms had electrical service that was provided through the R.E.A. or by other means. New demands for household electrical appliances spurred growth in home appliance manufacturing, and spawned the electrical and plumbing trades in rural communities. Electrical service also brought revolutionary new mediums of communication to rural farms, firms and households. Radio was followed by television, and the new streams of information narrowed the cultural, educational and commercial divide between urban and rural America. Rural electrification also contributed to the rapid growth of suburbs, and helped create a more integrated national market.

In 1994, Congress established the Rural Utilities Service (R.U.S.) as a federal agency within the United States Department of Agriculture, and it absorbed the R.E.A. and its responsibilities for rural electrification. Table 1 compares the share of the electric utility market for investor-owned companies, publicly owned companies, and rural cooperatives in the United States in 1998. Rural electric cooperatives still serve 11 percent of the nation's population and deliver 9 percent of the kilowatt hours sold. The data also indicate that rural markets continue to impose hardships to producers for costs and revenues. Cooperatives account for a much smaller portion of revenue per mile of wire (($7,873) than investor or publicly owned electrical utilities (($60,921 and $70,670,

respectively), and a greater portion of distribution plant investment per consumer.  To date, the R.E.A. and R.U.S. have organized nearly $57 billion in federally guaranteed low interest loans for the development of electric cooperatives.  The success of these programs, and the revolutionary outcomes, in both commercial sectors and in the organization of the home in rural America, is not unlike the outcomes to be expected from the penetration of the high-speed Internet into rural America.

**Table 1: Electric Utility Market Comparisons, United States, 1998**

|  | Investor Owned | Publicly Owned | Rural Cooperatives |
|---|---|---|---|
| Number of Organizations | 239 | 2009 | 930 |
| Customers, % of U.S. total | 74% | 15% | 11% |
| Revenues, % of U.S. total | 77% | 14% | 9% |
| Kilowatt hour sales, % of total | 75% | 15% | 9% |
| Number of consumers, per mile of line | 33 | 43 | 6 |
| Revenue per mile of line, in dollars | 60,921 | 70,670 | 7,873 |
| Distribution plant investment per consumer, in dollars | 1,890 | 1,870 | 2,352 |
| Assets, in $ billions | 606 | 126 | 70 |

Source: National Rural Electric Cooperative Association Strategic Analysis, March 1999, www.nreca.org/coops/elecoop3.html

**Internet Access in Rural America**

As was the case seventy years ago with rural electrification, private initiatives cannot deliver the infrastructure for affordable high-speed Internet access in sparsely populated rural areas of the United States today.  While many small cities and towns are beginning to reap the benefits of high-speed Internet access, much of rural America is being left behind as providing firms vie instead for more lucrative profits in densely populated cities and suburbs.  The problem for rural America is that plenty of firms are able to supply Internet content, but firms are generally unwilling to build the infrastructure necessary to deliver that content at a cost comparable to more densely populated locales.

Better telecommunications infrastructure is needed to deliver high-speed Internet service, and high-speed Internet service is needed to deliver band-consuming content.  Bandwidth measures information in flows of bits, which incorporate content and move over a given distance, in a period of time.

Broadband width, as compared to narrow, simply means that more information is transferred at vastly greater speed. With Internet content, the visual acronym WYSIWYG is still germane—what-you-see-is-what-you-get in terms of content that the infrastructure can load and carry. Content quality is determined solely by the telecommunications infrastructure that brings it to visual plates in households and workplaces. The dilemma of rural high-speed Internet access is not in the development of content, but in whether the telecommunications infrastructure can deliver content to rural homes and businesses.

Three core technologies furnish the non-trunk line infrastructure for Internet telecommunications: telephone lines, cable television wires, and wireless technologies for land-based antennas (fixed-wireless) and earth-orbiting satellites. While telephone wires have existed for over a century, fiber-optic cable has greatly expanded the carrying capacity of telephone wire from narrowband to broadband ranges. So, too, has the carrying capacity of cable television wire been enhanced—to the point where cable firms can offer analog video, digital video, Internet, and telephone services. Cable television originally brought higher content (in volume of channels) to many rural towns where "wireless" television broadcasting signals had been compromised by distance and terrain. Wireless telecommunications technologies for consumers emerged in the 1970s, when satellites brought the content of cable television to rural areas that cable companies failed to serve because they were too costly to wire, and cellular telephones brought a whole new architecture to voice communications. But rural America lags far behind in wireless telephone service, and many rural areas remain blacked out.

As modem users well know, telephone wires were the original carrying infrastructure for the Internet. Most Americans still access the Internet through a narrowband modem connection of 56k or less, over a dedicated or non-dedicated telephone line. Research firm Juniper Communications forecasts that 80% of users will remain connected to the Internet via standard modem in 2002.[14] Access to the Internet through telephone lines, at greater speeds, occurs through two broadband enhancements to the technology: fiber-optic cable and digital subscriber line service, or D.S.L. Using existing copper telephone lines, D.S.L. achieves information-transfer speeds that are ten times faster than typical phone modems. Along fiber-optic lines, D.S.L. offers the fastest connection attainable at a cost comparable to the best alternatives. Unlike cable Internet service, each customer has a dedicated line, which minimizes fluctuations in speed and provides a secure connection. But D.S.L. offers nothing to rural Americans because it is only available to households and firms located within 17,500 feet (a little over three miles) of a central telephone office, and the 30,000 offices nationwide are concentrated in densely populated areas.

The use of the original coaxial cable television wire for Internet telecommunications infrastructure also has considerable drawbacks. Coaxial cable provides one-way high-speed Internet service, since the wire was designed to bring in a few dozen channels of television, not to return signals to the cable provider.  Much of the national cable infrastructure was installed in the 1970s, and while the carrying capacity can bring Internet content into households and workplaces at high-speed, content sent out slows to speeds equivalent to a 56k modem or less.  Because of this limited capacity, cable companies are rushing to replace the coaxial infrastructure with a Hybrid Fiber-Optic Coaxial Cable System (H.F.C.) that compresses television, Internet, and telephone service into one line.  But technical problems compromise high-speed H.F.C. service.  If there is only one user of a cable connection in a neighborhood, the incoming speed over an H.F.C. network can approach 500 times faster than a 56k modem.  If, on the other hand, many neighborhood users simultaneously tap the same H.F.C. cable for Internet access, the incoming speed plunges to narrowband ranges.  An observer writes:  "If one of your cable-using neighbors starts running a Web server out of the home, you can imagine neighbors marching with torches and pitchforks." [15]

NxGen Data Research estimates that by 2003, cable will deliver high-speed Internet access to 8.4 million users, compared to 5 million D.S.L. users, 450,000 wireless broadband and 900,000 satellite users.[16]  Eighty percent of existing cable networks in the United States have been upgraded to H.F.C.[17]  AT&T, owner of TCI, is spending $15 billion alone to upgrade the TCI cable system.[18]  An observer recently noted that the "converging Internet, cable TV and telephone industries are spending billions of dollars to make broadband a reality—at an estimated construction cost of $400 to $500 a household, whether the broadband connection is through a cable system or a telephone line.  The effort represents the most extensive and expensive engineering project in residential communications since the cable TV industry started wiring neighborhoods in the 1960s."[19]  But the telecommunications infrastructure upgrade war is only being waged in American cities and suburbs.

It comes as no surprise that upgrades in cable and telephone infrastructure have occurred first in the most lucrative markets.  It is time to also acknowledge, from our national legacy with rural electrification and cable television, that the economies of connecting consumers separated by hundreds of feet or more will never justify building a state-of-the-art cable infrastructure in rural areas.  Were it not for the development and wide-diffusion of satellite television technologies in the last three decades, rural America would still be holding out for larger roof antennas to capture signals from a handful of distant television stations.  Wireless technologies that promise to deliver broadband Internet service consequently represent something akin to a Holy Grail to rural Americans hoping to share in the spoils of faster Internet access.  But how affordable and realistic are these technologies?

**Table 1:  Cost and Speed for Internet Service in Urban Areas**

| Service | Delivery | Monthly cost | Speed |
| --- | --- | --- | --- |
| Dial Up | Phone line | $20 | 56 kbps (or slower) |
| ISDN | Phone line | $50 - $130 | 128 kbps |
| Satellite | Phone line to wireless | $50 (for 100 hours) | 400 kbps (or slower) |
| Cable | Cable | $30 - $65 | up to 2.5  mbps |
| DSL | Phone line | $50 - $1200 | up to 8 mbps |
| Frame relay and T1/T3 | Dedicated fiber-optic | $300 - $3000 | up to 45 mbps |

Source:  PC World, "Six Routes to the Internet, with Speed and Cost," March 1999, p. 110.
Estimates are for urban areas, only.  A companion article, "Bandwidth on Demand," provides an
excellent overview of the competing technologies that deliver Internet access.

Cost and speed for infrastructure technologies that deliver Internet
content to urban areas are compared in Table 1.  We can assume from industry
unwillingness to invest that cable, D.S.L. and frame-relay are not cost-viable
technologies in sparsely populated rural areas.  If there is to be a new
infrastructure to deliver broadband Internet access to rural America it would
appear that wireless stands alone among alternatives in the market.  Data from
Table 1 indicates that the cost of wireless in urban areas is comparable to the
other choices.  Assuming that wireless service can be provided at comparable
cost, is the technology itself viable in rural areas?

Broadband wireless Internet access can be furnished either through a
land-based system of antennas (fixed wireless), directly from satellites, or from a
system combining the two.  But any land-based wireless Internet transmission
requires clear lines-of-sight between antennas, and signal obstructions are caused
not only by hills and valleys, but also by buildings, trees, and even precipitation.
A fixed wireless system that provides national coverage will require tens of
thousands of land-based antennas, and even then cannot guarantee
uninterrupted coverage.  Wireless Internet transmissions are also incompatible with other wireless networks, and marketplace posturing and rivalries among competing technologies has led to an impasse on an agreement for industry standards among firms.  Moreover, as anyone who has benefited from a baby monitor knows, security concerns plague wireless transmissions.  Industry observers acknowledge that wireless broadband is more viable for urban areas, where roof-mounted base-station antennas in a Wireless Local Loop (WLL) configuration can cover a radius that encompasses most of a downtown.  As for rural areas, fixed wireless broadband work best on prairies, deserts or other areas with uniformly flat topography.

Prospects are no better on the satellite-direct side of wireless technology, where communication remains a one-way medium between satellites and earth-based stations.  To access the Internet via satellite from a keyboard or mouse requires both a satellite connection and an open telephone line.  This limits the speed of Internet access to the bandwidth range of the telephone connection.  As Business Week dourly notes, satellite data must be pumped out over a "pokey standard phone line," and "neither fixed wireless nor satellite service is expected to compete heavily with D.S.L. and Cable Modems."[20]

Recent episodes with wireless satellite network communications further call the viability of the technology into question.  Between 1997 and 1999 NYSERNet conducted a trial of a wireless multi-channel multi-point distribution service (M.M.D.S.), providing Internet access to ten schools in the Rochester, New York area.  CAI Wireless Systems, Inc. (since merged with MCI WorldCom) used a high-speed 10-mbps satellite downlink connection for the experiment.  But the return, uplink path for this asymmetric service was a 28.8 kbps analog modem and a dedicated analog telephone line.  The NYSERNet final report on the experiment notes that in addition to the slow uplink, other obstacles included connection problems, weather interference, and government regulations.[21]  On a global scale, the decision by bankrupt Iridium, LLC to abandon and destroy its $5 billion worldwide telecommunications network of 88 satellites should give every champion of wireless Internet pause.[22]

**Regulatory Factors:  The Telecommunications Act of 1996 and the F.C.C.**

The Federal Communications Commission is required by provisions in the Telecommunications Act of 1996 to take regulatory measures to ensure comparable and affordable access to the Internet for all Americans.  In particular, Section 254.b.3 "Universal Service Principles: Access in Rural and High Cost Areas" of the Telecommunications Act of 1996 states:

"Consumers in all regions of the Nation, including low-income consumers and those in rural, insular, and high cost areas, should have access to telecommunications and information services, including interexchange services and advanced telecommunications and information services, that are reasonably comparable to those services provided in urban areas and that are available at rates that are reasonably comparable to rates charged for similar services in urban areas."[23]

To date, even with a change in Presidential Administrations, the F.C.C. maintains the position that broadband Internet is a nascent, "largely inchoate technology," and no remedial action is necessary to ensure universal access, or to require so-called carriers to directly contribute to universal access.[24] The F.C.C. stance is that the market will *eventually* provide universal service to all Americans, even though the costs make it unlikely that rural Americans will see D.S.L. or cable Internet architectures competing with wireless technologies.[*] From the standpoint of the F.C.C., rural American hopes for high-speed Internet access hinge on Teledesic, Starband, Hughes Network System's DirecPC, Wildblue, and SkyBridge. Of these, only Teledesic, a privately-held company owned by Microsoft's Bill Gates, cellular telephone pioneer Craig McCaw, Motorola, and Boeing, aims for complete global coverage. Teledesic proposes to launch 288 satellites in low earth orbit (L.E.O.) for two-way digital voice, video, and data transmissions. The system will cover 95% of the landmass surface and nearly 100% of the population of the planet at a cost of $9 billion. Teledesic originally expected to begin service in 1999, and periodically pushes the service start date (now 2005) further into the future.[25] But the technological pitfalls remain so deep that it will be a long time before wireless becomes the blanketing solution to broadband access in rural America.

Another major stumbling block to high-speed Internet access in rural America comes from a provision in the Telecommunications Act of 1996 concerning infrastructure. As was the case with the breakup of AT&T, the Act requires telecommunications firms owning infrastructure that delivers Internet content to share that infrastructure with other providers of content. Sec 259.b.4 "Infrastructure Sharing: Terms and Conditions of Regulations" states:

"The regulations prescribed by the Commission pursuant to this section shall ensure that such local exchange carrier makes such infrastructure, technology, information, facilities, or functions available to a qualifying carrier on just and reasonable terms and conditions that permit such qualifying carrier to fully benefit from the economies of scale and scope of such local exchange carrier, as determined in accordance with guidelines prescribed by the Commission in regulations issued pursuant to this section."[26]

---

[*] The single exception to the FCC hands-off approach is the E-Rate program, where a tax on long distance telephone service finances Internet connections for schools and public libraries. Enacted despite considerable Congressional skepticism, telephone companies pay a fee to the F.C.C., which uses the monies to promote point of service access. Telephone firms recover the fee through a tax that costs consumers around 30 to 40 cents per month. For 2000, the F.C.C. was authorized to collect $2.25 billion to further the initiative. Still, only 39% of rural classrooms, compared to 62% of urban and suburban classrooms are connected to the Internet. See Cheryl Rainford, "Bridging the Digital Divide" @griculture Online, July 14, 1999, www.agriculture.com/scgi/AgNews_tailNews_Asearch_listAgNews_html_41507.

The sharing requirement does not apply however to local cable television franchises, which creates a host of distorted incentives that breed comparative advantages and disadvantages in the market. The reigning incentive governing the high-speed Internet market is for firms to seek ownership and control of both infrastructure and content. This is why Time-Warner, with substantial investment and ownership in cable networks throughout the nation, paired with AmericaOnLine. While Time-Warner offered monopoly control over its cable infrastructure, it gained a first-mover advantage through the 20 million customers receiving their Internet content filtered through AOL. The marriage yielded strategic comparative advantages among media giants navigating the regulatory framework that is shaping the market. Other cable providers are extending the logic and consolidating monopoly control over urban and suburban markets through some rather novel arrangements. In 2000, Cablevision, the leading cable firm in metropolitan New York, traded 357,000 of its customers in the Boston-area to Media One for 125,500 New York-area Media One customers and $1.16 billion in cash and stock. Media One is a division of AT&T.[27] Meanwhile, no provider fights for territory or postures to build new infrastructure for high-speed Internet access in rural America.

The incentives to build rural high-speed Internet infrastructure are further dampened by provisions in the Telecommunications Act of 1996 that prohibit the Bell regional operating companies from directly competing in the market. Bell companies have no incentive to replace the existing telephone infrastructure in rural America with lines (perhaps even H.F.C. cable) capable of delivering broadband Internet content and telephone service. Freeing the Bell companies would encourage them to upgrade their rural telephone networks to provide affordable high-speed Internet access in places where D.S.L. cannot reach, cable companies have shown little past interest in wiring, and satellite and wireless await the future. Such improvements would acknowledge the technological convergence that has occurred in wiring technologies for cable television, telephone and Internet pipe. Removing the regulatory yoke would foster a competitive framework in the rural American Internet market with Bell companies and other telephone service providers upgrading their infrastructure as wireless service is being introduced.

The crux of the rural American dilemma is that there is currently no incentive to build land-based wiring infrastructure for broadband Internet access. Moreover, there are disincentives to build such networks, driven in part by market realities and in part by competitive restrictions in the Telecommunications Act of 1996 and the steadfast position of the F.C.C. not to intervene in the name of universal service.

**Voices Against Government Intervention**

Early voices opposed to any effort on the part of the federal government to address the lack of a high-speed Internet infrastructure in rural America were varied and considerable. In January 1998 the United States Internet Council published a focus group study of 60 interviews that it conducted with state legislators from 30 states. The question "what is the role of state government in bringing about the benefits of the Internet?" was met, in the words of the Council, by "silence…literal dead silence."[28] In comments on the merger of AOL and Time Warner, Bob Metcalfe, Founder of 3Com stated that the only thing "to worry about on the broadband Internet is preserving freedom of choice among competing alternatives. Worry about Big Business taking over the Internet if you want—AOL Time Warner, Microsoft and the telephone monopolies, for example—but be sure to keep an eye on Big Government."[29] In a press release issued at a broadband access hearing he conducted on April 11, 2000 Representative Tom Bliley, (R, Virginia) Chair of the House of Representatives Committee on Commerce, stated that the Internet

> provides a digital opportunity for all Americans. Some people have argued that services and products will only come to rural or urban centers if the federal government forces or mandates that it occur. I think that today's witnesses are only a small slice of the examples of companies and organizations that are trying to solve any perceived deployment disparity problem rather than look for a federal government program.

Nevertheless, a shift in thinking has occurred over the last two years among policy-makers who have contemplated the prospects for high-speed Internet access in rural America. In July 1999, the United States Commerce Department published "A Report on the Telecommunications and Information Technology Gap in America." The so-called Daley Report noted the presence of a 'digital divide' between suburban and urban areas and low-income urban and rural areas.[30] Senator John McCain, who ran a spirited campaign for President through the 2000 Republican Primary season, was a pioneer sponsor of legislation (S. 1043, 106th Congress, The Internet Regulatory Freedom Act of 1999) that sought improved broadband Internet access in rural America through regulatory reform which would free local telephone companies to compete with local cable television franchises. In March 2000 another early legislative initiative was proposed by a group of United States Senators with large rural constituencies—Dorgan, Daschle, Baucus, Johnson and Harkin. Their Rural

Broadband Enhancement Act (S. 2307, 106th Congress) would authorize the Rural Utilities Service of the United States Department of Agriculture to make loans and extend credit, from a $3 billion fund, to telecommunication carriers and affiliated companies to finance the deployment of broadband Internet service to rural communities. The loans, for 30 years at 2% interest, would have gone to firms extending D.S.L., cable, fixed wireless or satellite wireless infrastructure to any community of less than 20,000 inhabitants.

Legislative interest in broadband rural access has surged in the 107th Congress in 2001. Bills pending fall into three categories: opening the market to competition, tax credits for broadband build-out, and loan guarantees. Among the bills introduced, several propose remedies directed to these categories. H.R. 1542 focuses on deregulation, S. 88, S. 150, S. 246, H.R. 267, and H.R. 1415 would offer rural broadband deployment tax credits and S. 428, H.R. 1416 and H.R. 1697 propose grant and/or loan guarantees. Of these initiatives interest has centered on H.R. 267 (English), H.R. 1542 (Tauzin), and S. 88 (Rockefeller). H.R. 267 and S. 88, the Broadband Internet Access Act of 2001, would provide tax credits to firms investing in infrastructure build out in rural and low income areas. H.R 1542, the Internet Freedom and Deployment Act of 2001, would free Bell companies and incumbent telephone companies to provide broadband services in any market. Loan provisions appear to be the alternative least likely to gain legislative approval.

**Casting the Internet to Rural America**

The historical similarities and comparative analysis of rural electrification and high-speed Internet access suggests that the national goal of universal service is unlikely to be met in the near future. Still, there are lessons to be learned from the comparative historical policy-making assessment of these two episodes of market failure in rural America. Despite the objection that markets will eventually provide broadband services to rural America, a strong argument can be made for Federal government action to reverse the direction of a digital divide that grows by the day. There may be no greater contribution to economic development in the first decades of the 21st century than federal government leadership in creating an incentive-rich environment for constructing rural high-speed Internet infrastructure. Building upon a foundation of national success with the R.E.A., federal action should embody the best of progressive policy-making to nurture the market for high-speed Internet access in rural America.

A policy that creates incentives for firms willing to wire rural America for broadband Internet must and can guarantee that no individual or industry is comparatively advantaged by favoritism, add no regulatory yoke to the market, and impose no additional burdens upon taxpayers. The key component of such a policy is to foster the creation of infrastructure needed to carry bandwidth-rich Internet content to and from rural locales. Incentives, in the form of the long-term, low interest loans, need not discriminate among competing technologies. Even the F.C.C. has stated a position consistent with policies that

> "promote a competitive market by encouraging innovation, investment, and infrastructural build-out.  In so doing, government insures that innovative and cost-efficient services will be provided to consumers by a diversity of entities—or multiple pipes to the homes."[31]

The only significant hurdle to implementing an incentive-driven policy for rural broadband Internet access is that it has been difficult to determine, by conventional empirical measurement or spatial analysis, population distributions and densities that objectively identify what is meant by a 'rural' locality. Assuming that we can measure accurately, do we use, for example, 5,000 persons per municipality or less than 100 persons per square mile?  Moreover, localized population density measurements, and even definitions of the scale of S.M.S.A.s, are subject to endless debates on which communities to include or exclude. Questions with respect to the fairness of the criteria employed in making those determinations are especially vulnerable to criticisms of political influence and meddling.

Fortunately, a far more consistent and politically neutral definition of what is meant by 'rural' is provided by the current market for high-speed Internet infrastructure.  In this market-determined definition, federally-financed loans could be made to firms willing to bring broadband Internet infrastructure *to any geographic area that lacks access to cable television*.  The answer to the question of where to apply the federal incentives thus rests squarely in those areas that are without a cable television infrastructure, since similar disincentives in the market prevented cable television from being brought to these remote rural locales.  In this way, the rural market for high-speed Internet infrastructure, through a federally sponsored loan program, could be opened with equal access and competitive footing, to cable television companies, telephone companies or other providers of telecommunications services.  Likewise, it would be reasonable to extend federal loans to firms that propose only to build infrastructure, cued to a high bandwidth standard, with the intention of leasing it to other Internet content providers.

Loans designed to spur construction of broadband Internet infrastructure in rural America would differ, in one intriguing respect, from R.E.A. loans.  The broadband infrastructure loans provide incentives to firms to construct infrastructure, not to individuals to form cooperatives to construct infrastructure, as was the case with the R.E.A.  Direct loans to firms would negate fears of big- or micro-government management of Internet access in rural America. Moreover, since the low-cost loan incentive creates no shortage of competitors willing to build infrastructure, any fears concerning the monopolistic practices of publicly regulated utilities do not apply either.

Finally, with respect to federal funding, a loan program of this type would be both a fiscally prudent and progressive employment of federal monies, even in a changing environment where reconstruction, expenditures for military conflict, and the pressures of recession are likely to take precedence. With the remarkably low default rate on R.E.A. loans, the American public should expect the monies advanced for the construction of rural Internet infrastructure to be repaid and then recycled into new loans or returned to the federal treasury.

Lacking a high-speed rural Internet infrastructure, the consequences to the nation as a whole will increasingly mirror the plight of rural Americans without electricity in the decades prior to the Great Depression. The question is no longer whether to maintain the status quo in the market for broadband Internet access, since the digital divide worsens the status quo by the day. If sparsely populated regions continue to lag in high-speed Internet access, we tacitly accept diminished rural productivity, less substantial rural demands for commodities and services, lagging rural incomes, and widening income inequality. Those who choose to live in rural America for its environmental attractiveness, low crime, family-centered lifestyle, and democratic educational institutions will be comparatively disadvantaged over the next decades. When a federal effort to create incentives to invest in rural Internet infrastructure requires marginal investments that will eventually be repaid, and when a satisfactory formula to distribute those incentives to private sector investments can be devised, the absence of policy-making action to actively promote universal access is no longer justifiable.

---

**Endnotes**

1.  D.C. Brown, <u>Electricity for Rural America</u>, Westport, Connecticut: Greenwood Press, 1980, p. 5.

2.  Brown, <u>Electricity for Rural America</u>, pp. 6 and 29.

3.  Brown, <u>Electricity for Rural America</u>, p. 32.

4.  Morris L. Cooke, "National Plan for the Advancement of Rural Electrification Under Federal Leadership and Control with State and Local Cooperation and as a Wholly Public Enterprise," February, 1934, Cooke Papers, Box 230, Franklin Delano Roosevelt Presidential Library, Hyde Park, New York, p. 6. Cooke's "The Early Days of the Rural Electrification Idea, 1914-1936," <u>American Political Science Review</u>, Volume 42, June 1948 also provides a good overview of rural electrification during this period.

5.  Cooke, "National Plan for the Advancement of Rural Electrification Under Federal Leadership and Control with State and Local Cooperation and as a Wholly Public Enterprise," p. 8.

6.  Cooke, "National Plan for the Advancement of Rural Electrification Under Federal Leadership and Control with State and Local Cooperation and as a Wholly Public Enterprise," p. 9.

7.  Cooke, "National Plan for the Advancement of Rural Electrification Under Federal Leadership and Control with State and Local Cooperation and as a Wholly Public Enterprise," p.11.

8.  Brown, <u>Electricity for Rural America</u>, p. 68.

9.  Brown, <u>Electricity for Rural America</u>, p. 69.

10. Paul L. Joskow and Richard Schmalensee, <u>Markets for Power:  An Analysis of Electric Utility Deregulation</u>, Cambridge, Massachusetts:  M.I.T. Press 1983, p. 17.

11. Brown, <u>Electricity for Rural America</u>, p. 41).

12. Sam H. Schurr, Calvin C. Burwell, Warren D.Devine Jr., and Sidney Sonenblum, <u>Electricity in the American Economy:  Agent of Technological Progress</u>, Contributions in Economics and Economic History, Number 117, Westport, Connecticut: Greenwood Press, 1990, p. 234.

13.  Brown, <u>Electricity for Rural America</u>, p.114.

14.  Harry McCracken, "Bandwidth on Demand,"  <u>PC World</u>, March 1999, p.188.  For an account of the struggles to get more speed see Ron Brownstein, "Rules of the Road Required to Maintain Top Speeds on the Information Highway," <u>Los Angeles Times</u>, November 29, 1998.

15.  Peter Lewis, "High-Bandwidth Web Access Has Its Speed Bumps." <u>New York Times</u>, November 11, 1999, p. G10.

16.  Roger O. Crockett. and Andy Reinhardt, "Faster, Faster, Faster: Companies are Giving Net Users what they Demand:  Speed." <u>Business Week</u>, October 18, 1999, p. 194.

17.  Federal Communications Commission Staff Report, <u>Broadband Today</u>, Prepared for William Kennard, Chairman, Federal Communications Commission, October 1999, 26.

18.  Crockett and Reinhardt, "Faster, Faster, Faster," <u>Business Week</u>, October 18, 1999, p. 192.

19.  Seth Schiesel, "Broadband:  How Broadly?  How Soon?" <u>New York Times</u>, January 17, 2000, p. C1.

20.  Roger O. Crockett. and Andy Reinhardt, "Where to Find Warp Speed," <u>Business Week</u>, October 18, 1999, 194.

21.  Robin L. Jones, "NYSERNet-CAI Wireless Internet Connectivity Trail," Final Report, October 29, 1999 www.nysernet.org/wireless.html.

22.  <u>New York Times</u>, "Lacking Buyer, Iridium is Set to Shut Down," March 18, 2000, p. C1.

23.  Federal Communications Commission, <u>www.fcc.gov/telecom.html#text</u>.  See also Patricia Aufderheide, <u>Communications Policy and the Public Interest:  The Telecommunications Act of 1996</u>, New York:  Guilford Press, 1999.

24.  Federal Communications Commission Staff Report, <u>Broadband Today</u>, p. 6.  See also F.C.C. Chairman William E. Kennard, "How to End the World Wide Wait," <u>Wall Street Journal</u>, Op-ed, August 24, 1999 and the comments of F.C.C. Commissioner Susan Ness, "Remarks Before the Annual Meeting of the National Telephone Cooperative Association," February 14, 2000.

25.  Federal Communications Commission Staff Report, <u>Broadband Today</u>, p. 30; www.teledesic.com/about/about.htm.

26.  Federal Communications Commission, <u>www.fcc.gov/telecom.html#text</u>.

27.  Jason Blair, "Cablevision Makes Deal to Add to Its New York Customers," <u>New York Times</u>, April 25, 2000 p. B7.

28.  United States Internet Council, <u>Internet Survey of State Legislators</u>, January 1999 <u>www.usic.org/survey.html</u>.

29.  <u>New York Times</u>. "At the AOL-Time Warner Wedding, Visions of Better and Worse," January 16, 2000 C4.

30. United States Department of Commerce,  <u>A Report on the Telecommunications and Information Technology Gap in America,  National Telecommunications and Information Administration</u>, July 1999, William M. Daley, Secretary of Commerce, Washington: U.S.G.P.O. 1999.  For a rural perspective see Mary Ann Barton and KevinWilcox, "Broadband Rollout in Rural Areas is No Party Line," <u>County News</u>, National Association of Counties, August 23, 1999.  Other perspectives are found in Robert D. Atkinson, "Social Policy in a Competitive Marketplace: The Need for Telephone Universal Service Reform," Washington:  The Democratic Leadership Council and The Progressive Policy Institute, January 2000  <u>www.dlcppi.org/texts/tech/universal.htm</u> and Matt Robinson, <u>A 21st Century Internet for All Americans</u>, Report Prepared for iAdvance, December, 1999.

31. Federal Communications Commission Staff Report, <u>Broadband Today</u>, p. 41.